# Information Delivery System through Bluetooth in Ubiquitous Networks

**D.Asha Devi**,
Associate Professor of ECE, INTELL Engineering College, Anantapur – 515004.
Email : ashadevi.d@rediff.com

**M.Suresh Babu**
Principal , INTEL Institute of Science, Anantapur-515004,
EMail : principaliis@rediff.com

**V.L.Pavani**,
Principal, Beasant Institute of Technology & Science, Anantapur – 515004.
Email : pavani_v_l@yahoo.co.in

**Dr.N.Geethanjali**
Associate Professor of Comp. Applns, Sri Krishnadevaraya University, Anantapur-515053

-------------------------------------------------------ABSTRACT-------------------------------------------------------
Ubiquitous and pervasive computing (UPC) is a popular paradigm whose purpose is to emerge computers into the real world, to serve humans where the ubiquitous network is the underneath infrastructure. In order to provide ubiquitous services (u-Service) which deliver useful information to service users without human intervention, this paper implements a proactive information delivery system using Bluetooth technology. Bluetooth is a lowpowered networking service that supports several protocol profiles, most importantly file transfer.Combined together, ubiquitous computing and Bluetooth have the potential to furnish ubiquitous solutions (u-Solutions) that are efficient, employ simplified design characteristics, and collaboratively perform functions they are otherwise not capable. Thus, this paper first addresses the current Bluetooth technology. Then, it suggests and develops the proactive information delivery system utilizing Bluetooth and ubiquitous computing network concepts. The proactive information delivery system can be used in many ubiquitous applications such as ubiquitous commerce (u-Commerce) and ubiquitous education (u- Education)

## 1. Introduction

Since Weiser envisioned ubiquitous and invisible computing as an opposite concept of virtual reality, the ubiquitous and pervasive computing (UPC) has gained popularity rapidly due to its convenient accessibility from a user's prospective [11]. In fact, whereas virtual reality makes people enter a computer-generated world, UPC makes computers come out to the real world to serve humans. Thus, the term ubiquitous computing is abstractly defined as the integration of computers into the environment while maintaining the illusion of invisibility to the user [11].

To provide customized services to users without letting them login to computers, it is essential to include intelligent and proactive service activators which initiate the services for users without human intervention based on requests which are preset by the users. For instance, in ubiquitous education (ueducation), a service activator such as a proactive information deliverer can automatically initiate the transfer of class notes to Bluetooth enabled devices such as a Personal Digital Assistant (PDA), cellular phone, or computer as a registered student enters the classroom. In ubiquitous commerce (u-commerce), a proactive information deliverer also can transfer a customized catalog based on the customer's preset preferences upon entering the store. The system also can be used in the following example. Using a low powered wireless transmission radio such as Bluetooth; an auto mechanic may receive diagnostic data from a vehicle as it enters the garage, without the need for interaction; or a digital camera that sends photographs to a Bluetooth enabled device; all of which are probable scenarios; controlled wholly within a ubiquitous network. Similar activities using Bluetooth anonymously already exist in the entertainment industry, though research is extending into health care, the home and public meeting arenas. These ideas provide a basis for determining barriers and demand levels for the availability of such technologies thereby, providing a viable foundation for a proactive information delivery system in this paper. The system hones an abstract model and is designed to be portable and extensible to multiple platforms and environments as a result of the marriage between ubiquity and Bluetooth.

The rest of this paper is organized as follows. In Section II and III, motivation and existing research are addressed. A proposed idea, a development, and a demonstration are illustrated in Section IV, V and VI respectively. Then, results, further research, and concluding remark follow.

## 2. Motivation

The proactive information delivery (PID) system is developed to achieve two objectives. First of all, it is developed to provide intelligent and proactive ubiquitous services which automatically initiate customized services for a mobile user based on the user's taste and preference whenever the user enters a certain area such as a classroom or a store. Secondly, it also aims to save natural and financial resources such as papers and marketing costs. For example, a simulated classroom with 18 students receives paper distributions containing course content at 3 pages per week or 54 total pages per week. A typical 17 week semester would translate to 918 individual pages distributed. Therefore, based on assumption, a college campus that employs 800 instructors, where ¼ are teaching four courses or more per week and using 1836 pages per semester for two of the four courses, then approximately 367,200 pages are used, which equates to 708 reams of paper or 44 trees, where 8,300 pages =16 reams = 1 tree. The proposed idea is to prove and reduce the amount of paper printed in a single course for one semester using ubiquitous computing and Bluetooth.

## 3. Existing research

Finding a viable solution requires investigation of existing procedures in order to generate a comparative illustration. This section describes possible current scenarios and explains the benefits a ubiquitous controlled file transfer system. In this paper, the PID module is called a client since it also plays a role as a client for a service discovery and composition mechanism in a ubiquitous network environment. The Bluetooth enabled device such as a PDA, cellular phone, or computer is called a server in this paper since it also plays a role as a service module provider for a service discovery and composition mechanism. A viable solution is through the use of Bluetooth and ubiquitous computing. Bluetooth is a low-powered radio communications technology that is used in cell phones, laptops, PDAs and other consumer products on the market. The Bluetooth standard defines profiles which may offer services such as file transfer, printing, or synchronous/ asynchronous transmissions.

Therefore, the proposition, in order to maintain invisibility, is to develop a software package that will reside on a client machine and automatically search for Bluetooth devices or servers; when a device is in range of the signal radius, the client will query the server to establish that it uses the File Transfer Protocol (FTP) service and then transfer a file, using that service. There are several issues to contend with using this solution- authentication, authorization, updates (broadcast) and failures, which will be addressed in further research.

## 4. Bluetooth Device Stack

Bluetooth is a low-powered wireless technology that utilizes several different protocols for handling a plethora of processes. Bluetooth has encryption capabilities, file transfer, synchronous and asynchronous data transfer, and many other services available for use. And recently, with the release of Bluetooth version 2.1 + EDR (Extended Data Rate), transfer rates are capable of reaching 3.0Mbps [2] and is envisioned to become faster. The Bluetooth specification describes several methods of interfacing with the Bluetooth protocol stack through application development [2]. However, generating an extension is beyond the scope of this project. There are several third party interfaces that exist, concluding that designing a solution should be well defined prior to implementation. This section explains the multitude of stack interfaces, their benefits and absence thereof.

### 4.1. Python

The Python programming language uses the PyBluez extension for application development and is targeted for the Windows and Linux environments. PyBluez will not work with the Broadcom stack or with the L2CAP and HCI layers of the Bluetooth protocol stack [9].

### 4.2. VC++

The C programming language can be used in both the Linux and Windows environments. BlueZ is the extension for Linux and Microsoft has their own extension which has been integrated into the Windows OS distribution, beginning with XP. As previously mentioned, the different hardware stacks (Microsoft and Broadcom) cause difficulty for developers; the lack for a standard leaves many applications in a hope it-works state.

### 4.3. Java

The Java Community Process (JCP) developed an Application Programming Interface (API) in early 2002 as part of a competitive movement. The released package is known as JSR-82 (Java Specification Request). The JSR-82 API supports the RFCOMM and L2CAP transport layers [7]. There are several stack extensions available for the Bluetooth protocol stack, notably: ElectricBlue, BlueSim, and BlueCove. All three utilize the JSR-82 API but only one is free- BlueCove.

## 5. Development

The setup procedures are quick and simple; designing a command line interface (CLI) or graphical user interface (GUI) can be released in a short amount of time. The algorithms used are similarly simple and require only basic knowledge of Object-Oriented programming.

### 5.1. Setup

For this experiment the Eclipse integrated development environment (IDE) is used for code writing and BlueJ is used for debugging purposes. The application requirements are that the JSR-82, JSR-75 and BlueCove jar files be successfully added to the project's classpath. The JSR-82 and JSR-75 files are packaged in Sun's Wireless Toolkit 2.5.2 for CLDC (Connected Limited Device Configuration) package [10]. BlueCove is available through Google's code library. Seven classes were created for this project: A main driver, controller, application, utilities, file transfer, input output, and a device class. The MainDriver class creates a secondary thread so that the controller class may be run outside the main thread (Fig. 1). This is vital, along with creating additional threads inside the controller and application class. It may take up to 10 seconds to successfully query for connectable devices. Therefore, in order to retrieve all available devices, within range, it is necessary to halt executing threads while a discovery process thread continues.

```
Public static void main(String[ ] args)
{
   EvenQueue.invokeLater(new Runnable() {
      Public Void run() {
         Controller ct = Controller.getInstance();
         Ct.startApp();
      }
   } );
} // end main
```
**Figure 1 : Classes - Main**

The Controller class manages program operation. It creates all required objects and processes each method in a stepped procedure (Fig. 2). The IO class is used in the controller class to print application progress to the console.

```
Public void startApp( ) |
//display opening
10.pause("********Ubiqutious computing through
Bluetooth**********\* * A Java Application*);
//Step 1. Check if ccleint power is on
10.pause("Step 1. is the power on ?");
10.pause (this.app.getlocalDevicepowerState||);
//******************************************

//step 2.display client friendly name
10.pause("step 2. Local device name (this client);*);
10.pause (this.app.getLocalDeviceName()|;
//*********************************

// step 3. display client address
10. pause ("step 3. Local device address(thin client):");
10.pause(this.app.getLocalDeviceID(1);
//*********************************
```
**Figure 2. Classes : Controller**

The Application class is the meat of the program. The class initiates device discovery and service search. The application class implements the Discovery Listener class, which is part of the JSR-82 specification (Fig. 3).

```
public class Application implements DiscoveryListener {

   private LocalDevice localDevice;
   private DiscoveryAgent agent;

   public Application() throws BluetoothStateException {
      localDevice = LocalDevice.getLocalDevice();
      agent = localDevice.getDiscoveryAgent();
   }

   public void inquiry() throws BluetoothStateException {
      agent.startInquiry(DiscoveryAgent.GIAC, this);
   }

   public void deviceDiscovered(RemoteDevice device, DeviceClass devClass) {
      IO.display("Device discovered");
   }

   public void servicesDiscovered(int transID, ServiceRecord[] servRecord) {
      IO.display("Service discovered");
   }

   public void serviceSearchCompleted(int transID, int responseCode) {
      IO.display("Service search completed");
   }

   public void inquiryCompleted(int discType) {
      IO.display("Inquiry completed");
   }
}
```
**Figure 3. Classes : Application**

The Utilities class provides static, generic operations, such as traversing through data objects, parsing text, length and Boolean calls. There is another method for determining the MAC (Media Access Control) ID of a Bluetooth device as it is provided in the connection URL; more on this later. The MAC ID from the connection URL is valuable when making comparisons against device objects. The FileTransfer class simply creates a new File object, based on a file path, and then converts it into a

```
Step 4. Query for local devices.

Starting device inquiry…
Device Inquiry Complete; 5 devices were discovered
```
**Figure 4. Testing**

byte array. The FileTransfer class uses the Connection class from the javax.microedition.io package and the Client Session, HeaderSet, and Operation classes from the javax.obex class. These are necessary to create a session, set the file metadata and use the push operation, respectively, to send the byte array. The IO class is a static class that handles console input and outputs. There are several overloaded methods for ease of programming and a BufferedReader object that permits using the "enter" key for console input instead of passing an actual value into the console.

The BTDevice class is a data object that is created when a device is discovered. The class maintains generic getters and setters for the MAC ID, displayable name of the device, the connection URL (Uniform Resource Locator) and various other values. The basis for the BTDevice class is making connections and simplifying sending data, for example the MAC ID and the connection URL.

## 6. Testing

Although the PID system can be used in various platforms such as u-Education, u-Commerce, and u-Health, this paper tests the system in a classroom platform. Testing the proposed solution required devices that would typically be used in a classroom which incorporate the file transfer protocol. For example, cell phones, laptops, PDAs, or printer- in the case when a student does not have a Bluetooth device. The only instance that required modification was testing on a laptop and printer, in which case a D-Link Bluetooth dongle (model # DBT-120) was added to the USB port of the laptop and a Belkin Bluetooth Wireless Printer Adapter (model# F8T031) was used on the printer. Execution for the demonstration is a stepped procedure, based on method call in the controller class and a method call in the IO class, that performs the next operation when the enter key is pressed. Initial startup of the demo shows that BlueCove has been initialized and is running properly (Fig. 4).

```
BlueDeve version 2.0.1 on winsock
********Ubiquitous Computing through Bluetooth*****
              A Java Application
```

**Figure 4 : Testing – Step 0**

The first step ensures that the client's Bluetooth stack is powered up. Intuitively the host is the server however, in Bluetooth the client is the host device, as it houses the desired resource to transmit. The server is the device that utilizes a service the client is requesting for use. The second and third steps are not equired. The second step merely queries the client's friendly name while the third step requests the local device for its Media Access Control (MAC) ID. The fourth step is where actual device discovery begins (Fig. 5). Device discovery takes approximately sixteen seconds to perform. The method call is run in a separate thread from the controller thread as well as all subsequent method calls.

```
Step 4. Query for local devices

Starting device inquiry….
Device Inquiry Complete; 5 devices were discovered
```

**Figure 5. Testing – Step 4**

When a device is discovered it is added to a collection and displayed to the console (Fig. 6).

```
Step 5. The following devices were discovered :

Device 1.Dell BT keyboard – MAC id: 000761571B
Device 2. Dell BT Mouse – MAC id : 0007616110B1
Device 3. Interlink VP6600 Media Remote Control –MAC id:00150810BE
Device 4. DELL BH200 –MAC id:00164410697A
Device 5. EB-LAPTOP-MAC id:00179A235EDD
```

**Figure 6. Testing – Step 5**

Step six involves querying the devices found, from the discovery process in step 5, for their services offered (Fig. 7). For demonstration purposes all services are listed but upon initiation of a file transfer operation, only those devices that offer the FTP service are listed.

```
Step 6. Query all the devices for services offered.

Starting service inquiry...
Interlink VP6600 Media Remote Control
DELL BH200
EB-LAPTOP-D400
Service Query Complete;  3 devices have services:
Interlink VP6600 Media Remote Control - Number services: 1
DELL BH200 - Number services: 4
EB-LAPTOP-D400 - Number services: 7
```

**Figure 7. Testing – Step 6**

Step seven lists all the services a device offers (Fig. 8). The listing format is: service id, service name, connection URL. The connection URL is required for a given service in order to connect to that service.

**Figure 8. Testing – Step 7**

Step eight involves actual file transmission. The demonstration required that a valid file path be entered into the console, although it is possible to utilize a GUI interface (Fig. 9).

```
Step 8.Transfer file to a device.

Please type the complete file path for the desired file to be transferred.
Default path is: c1\icpi.txt
NOTE: error checking in 100 performed. Typing an incorrect path will result in application failure.
```

**Figure 9. Execution – Step 8**

During live testing the application searched for any file within a selected directory and transferred the file (the chosen directory only contained a single file) to the device accordingly. The first test subject was a Dell D400 Latitude notebook with attached D-Link Bluetooth dongle. A successful connection was made and the file was transmitted at a range of twenty feet.

The second test subject was an LG PM-325 cellular phone. Again, a successful connection was made and the file was transmitted with a range of twenty feet.

Live testing was conducted in a classroom of twelve students where seven Bluetooth enabled devices, using the file transfer protocol, were present. All seven devices successfully connected and received the document. No other testing was performed. The application was modified slightly to simulate a ubiquitous action: the majority of the demonstration steps were taken out so that only device discovery and service discovery were present; member verification (defined by a prior/initial link setup) existed; and device discovery occurred at intervals for an eight minute duration that overlapped the course start time (Fig. 10). Control subjects were used during live testing which consisted of two devices: one of which was not defined as a member of the course and the second as dummy device.

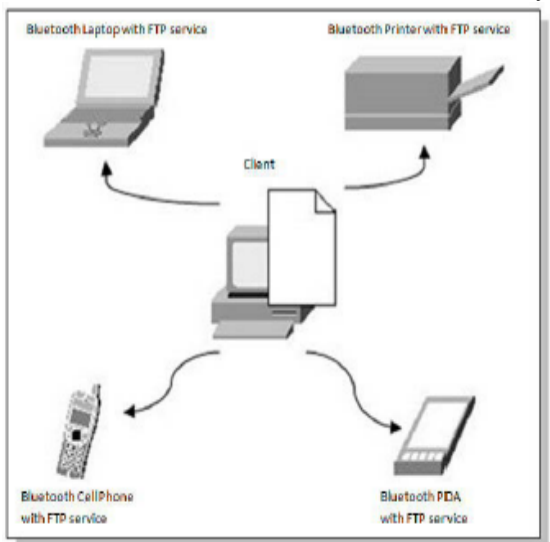

Figure 10.  Live Testing

## 7. Results

The results of the live test conclusively proved that file transfer to devices in a classroom was possible and feasible. The students who received an electronic version of the file in the live test were not given a printed copy. Four of four students reported no adverse affects on learning impedance; and all agreed that electronic collection was the preferred method over hard copy handout. The main reason being the ability to make corrections if necessary, or receive updated versions from the instructor. Adding ubiquitous computing to the project was simply a matter of enclosing the test program in a while loop; exiting the loop when either the file has been sent to all connected devices that are valid for that classroom or synchronized with a timer. The timer function would be more of an administrative purpose than pragmatic benefit. For example, an instructor may chose to not hand out files to students who are late to class.

## 8. Further research

The PID application did not cover some issues, though not problematic. There are, however, instances where a solution such as this may present itself as a realistic approach in public industry such as retail and maintenance. Some aspects were not covered for this simulation, for example: authentication, authorization, encryption, broadcast and failures. A complete solution would require that a form of authentication is performed to verify that a Bluetooth enabled device does indeed belong to the class/course it is attempting to serve the transfer protocol to additionally, the situation may arise where encryption is required, as this topic did not investigate the effects of encryption on the server device. However, it may be possible to implement such a service as the Bluetooth stack does support high encryption.

Another subject that was ignored was a broadcast service for updates. The update performed during testing was linear in approach. A true broadcast would require establishing a piconet with the server and clients. The drawback is that the Bluetooth specification states that in any piconet only one master and seven slaves may exist [2]. Although several issues were not addressed the application still proved useful. The main benefit is its portability. Considering that the language chosen was in Java, this scenario may be ported to several other cases. For example, a professor may conduct a Bluetooth exam, such that each student submits the exam before the end of the class. The difference between this and an Ethernet or web based exam scenario is the cost to the institution- regarding maintenance hardware purchasing, and security; overall, a low-cost solution. Thus, completing the design statement- find an alternative way of distributing files that would be cost-effective and save paper.

## 9. Conclusion

Ubiquitous computing has the ability to simplify, minimize, or altogether reduce human interventions currently required of computer system architectures. Bluetooth is a low-powered networking service that supports several protocol profiles, most importantly file transfer. Combined together, ubiquitous computing and Bluetooth have the potential to furnish solutions that are cost efficient, employ simplified design characteristics, and collaboratively perform functions they are otherwise not capable of. Thus, this paper first addressed the current Bluetooth technology. Then, it suggested and developed a proactive information delivery system with Bluetooth technology and ubiquitous computing network concepts.

## Author's Biography

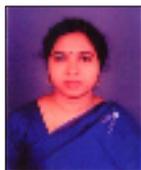
**D.Asha Devi** received her AMIETE from Institute of Electronics and Telecommunication Engineers in 2000. M.Tech (Digital Systems and Computer Electronics) degree in 2005 from JNTU College of Engineering, Anantapur(Autonomous). She served JNTU College of Engineering as lecturer in electronics from 2001 to 2004, later she moved to Intell Engineering College as Associate Professor. She is life member of IETE & ISTE. Many students have completed their B.Tech dissertation under her guidance and three students have registered for M.Phil. Her research interests are in VLSI, Microwave, Digital systems and embedded systems.

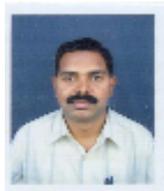
**M.Suresh Babu**, born on 27th December 1970 obtained his MCA from Osmania University and M.Phil Degree in Computer Science from Bharathiar Univeristy. At Present he is doing research in Computer Science on Inductive Data Mining. He served Intel Institute of Science, Anantapur in various capacities like Lecturer, Head of the Department and at present is working as Principal. He has organized several workshops and training Programmes in the field of Computer Science and attended a number of Workshops and seminars. He is a life member of ISTE and Science & Society & ISCA. He is nominated as State Convener, Education Sub Committee, A.P.Jana Vignana Vedika.

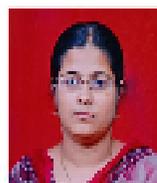
**V.L.Pavani**, received her MCA from Sri Krishnadevaraya University, Anantapur. She joined Beasant Institute of Technology & Science as Lecturer in 2002 and promoted as Principal in 2007. She has completed M.Phil from Bharathiar Univeristy in 2006. She has contributed more than 10 papers in various national and international conferences. She organized one National level workshop and acted as Coordinator for Faculty Development Program. She is doing research in Computer Science. Her area of interest is Algorithms, Data mining, Software Engineering.

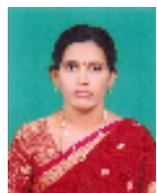
**Dr.N.Geethanjali**, has obtained Master of Science Degree from Sri Venkateswara University, Tirupathi. She has obtained PhD in 2004 from Sri Venkateswara University. She is working as Associate Professor in Department of Computer Science & Applications . She has more than 18 years of teaching experience for both UG and PG Courses.Her area of interest are Data mining, Data Communications, Artificial Intelligence, Cryptography, Network Security, Programming Languages.